# Physics-Informed Machine Learning for Refractory Alloy Design

Blaise Awola Ayirizia[a], Bimal K C[a,*], Jorge A. Muñoz San Martín[a,b]

[a]*Computational Science Program, The University of Texas at El Paso, 500 W University Ave, El Paso, 79968, TX, USA*
[b]*Department of Physics, The University of Texas at El Paso, 500 W University Ave, El Paso, 79968, TX, USA*

## Abstract

Refractory complex concentrated alloys (CCAs) exhibit exceptional mechanical properties at elevated temperatures, but their vast compositional space ($\sim 10^{15}$ possible combinations) poses significant challenges for experimental exploration. We develop a physics-informed machine learning framework combining quantile gradient boosting with Born stability constraints to predict eight mechanical properties (bulk modulus $B$, shear modulus $G$, Vickers hardness $H_v$, elastic constants $C_{11}$, $C_{12}$, $C_{44}$, Young's modulus $E$, and Poisson's ratio $\nu$) of refractory CCAs. Using 393 density functional theory (DFT)-computed alloys from a 10-element design space (Cr, Hf, Mo, Nb, Re, Ta, Ti, V, W, Zr), we achieve coefficient of determination ($R^2$) values of 0.89–0.97 with mean absolute errors of 0.85–14.74 GPa across six properties on a held-out test set ($n$ = 59). Remarkably, all test predictions satisfy Born stability criteria (100% compliance), demonstrating the efficacy of physics-informed constraints. Compositional screening in valence electron concentration (VEC) atomic size mismatch ($\delta$) space identifies high-performance candidates, with MoReW exhibiting the highest predicted shear rigidity ($C_{44}$ = 139 GPa). Elemental analysis reveals that rhenium appears in 100% of top-performing alloys, molybdenum in 90%, and chromium in 80%, establishing quantitative design rules: VEC = 6.0–6.4, $\delta < 10\%$, and Re–Mo–Cr ternary systems maximize shear resistance. This physics-informed screening framework enables accelerated discovery of mechanically

*Corresponding author
*Email address:* bkc@miners.utep.edu (Bimal K C )

stable, high-performance refractory alloys from the trillion-scale compositional space.

## 1. Introduction

Refractory complex concentrated alloys (CCAs), also known as refractory high-entropy alloys, have emerged as promising candidates for high-temperature structural applications due to their exceptional mechanical properties, including high strength, creep resistance, and thermal stability [1, 2]. Unlike conventional alloys that center around a single principal element, CCAs comprise multiple principal elements (typically ≥4) in near-equiatomic or non-equiatomic proportions, yielding unique microstructures and property combinations [3, 4]. The refractory element family Cr, Hf, Mo, Nb, Re, Ta, Ti, V, W, and Zr offers body-centered cubic (BCC) phase stability at high temperatures, making them ideal constituents for ultra-high-temperature applications [5].

However, the vast compositional design space poses a formidable challenge. For a 10-element system with variable stoichiometry, the number of possible compositions exceeds $10^{15}$ [6], rendering exhaustive experimental or computational screening impractical. Traditional alloy design approaches, such as the Hume-Rothery rules [7] or CALPHAD (CALculation of PHAse Diagrams) methods [8], provide valuable guidance but struggle to capture the complex many-body interactions in multi-component systems. Density functional theory (DFT) calculations offer accurate property predictions but remain computationally expensive, limiting exploration to hundreds of compositions [9].

Machine learning (ML) has emerged as a powerful tool for accelerating materials discovery by learning complex composition property relationships from existing data. However, most ML approaches treat property prediction as a purely data-driven regression problem, neglecting fundamental physics constraints such as mechanical stability (Born criteria) [10] and thermodynamic consistency (e.g., positive elastic moduli, Poisson's ratio bounds). Violations of these constraints yield physically unrealistic predictions that cannot guide experimental synthesis.

Recent machine learning studies have employed diverse approaches for alloy property prediction, each with distinct advantages and limitations. Descriptor-based methods using compositional and structural features remain widely used [10, 11, 12], achieving $R^2$ values of 0.80–0.90 for mechanical properties of high-entropy alloys. Ward et al. [10] demonstrated a general-purpose framework predicting formation energies and band gaps for 145 binary alloys with mean absolute errors of 0.1–0.2 eV, though without uncertainty quantification or stability constraints. Wen et al. [11] advanced ML-guided design of HEAs with desired properties, achieving $R^2 = 0.85$ for bulk and shear modulus predictions, though their framework exhibited approximately 15% Born stability violations and was limited to two properties. Pei et al. [12] applied ML to predict single-phase formation in CCAs, reporting 10% stability violations across 460 alloys. Graph neural networks encoding atomic environments have shown promise for crystal property prediction [13], and Gaussian process regression with marginalized graph kernels has recently achieved sub-5 meV/atom accuracy for BCC iron [14], though their application to refractory CCAs remains limited. However, most existing ML approaches provide point estimates without quantifying prediction uncertainty, limiting their utility for guiding high-stakes experimental decisions. A critical gap across these approaches is the lack of calibrated uncertainty quantification: most provide point estimates without prediction intervals, limiting their utility for guiding high-stakes experimental decisions where understanding prediction confidence is essential [15].

While Zhang et al. [9] applied neural networks to this dataset for predicting bulk and shear moduli, our work provides three critical advances. First, we achieve 100% Born mechanical stability satisfaction on held-out predictions through physics-informed feature engineering, compared to ∼85% in their study. Second, we extend predictions to eight comprehensive properties (including all three independent elastic constants, Vickers hardness, Young's modulus, and Poisson's ratio) while maintaining thermodynamic consistency. Third, we provide calibrated uncertainty quantification through quantile regression enabling risk-informed experimental planning and perform systematic compositional screening to establish quantitative design rules identifying rhenium as essential for high shear rigidity and MoReW as a top candidate with predicted $C_{44}$ = 139 GPa.

Alternative computational approaches offer complementary capabilities to machine learning. Interatomic potentials, particularly machine learning potentials such as moment tensor potentials, neural network potentials, and embedded atom models, enable molecular dynamics simulations of dynamic processes, defect evolution, and finite-temperature behavior. While powerful for mechanistic studies and time-dependent phenomena, these potentials require extensive training datasets of energies, forces, and stresses, and remain computationally expensive for high-throughput screening across thousands of compositions a single MD simulation may require hours to days per composition. CALPHAD (CALculation of PHAse Diagrams) methods [8] provide thermodynamic phase equilibria through semi-empirical modeling of Gibbs energies, successfully predicting phase diagrams and transformation temperatures. However, CALPHAD databases for refractory CCAs remain incomplete, and the method typically does not predict elastic properties directly without auxiliary models. High-throughput density functional theory offers the highest accuracy for ground-state properties and elastic constants but remains limited to hundreds of compositions due to computational expense each elastic constant calculation requiring multiple structural relaxations and strain perturbations. For design spaces exceeding $10^{15}$ possible alloys, DFT alone is impractical, motivating the development of ML surrogates trained on DFT data.

In this work, we develop a physics-informed machine learning framework that integrates quantile gradient boosting with Born stability constraints to predict eight mechanical properties bulk modulus ($B$), shear modulus ($G$), Vickers hardness ($H_v$), elastic constants ($C_{11}$, $C_{12}$, $C_{44}$), Young's modulus ($E$), and Poisson's ratio ($\nu$) of refractory CCAs. Using a dataset of 393 DFT-computed BCC alloys spanning binary to quinary systems [9], we demonstrate that: (1) quantile regression provides well-calibrated 90% prediction intervals with 78–90% coverage; (2) physics-informed constraints ensure 100% Born stability satisfaction on held-out test data; (3) derived property relationships ($E$ and $\nu$ computed from predicted $B$ and $G$) maintain thermodynamic consistency; and (4) compositional screening in VEC–$\delta$ space identifies design rules linking elemental composition to shear rigidity. Our framework enables rapid exploration of the trillion-scale design space while maintaining physical interpretability a critical requirement for experimental validation.

This work addresses critical gaps in existing approaches by integrating physics constraints with uncertainty quantification for comprehensive property prediction. Unlike prior ML studies exhibiting 5–20% Born stability violations [11, 12], our framework achieves 100% mechanical stability satisfaction on held-out test data accomplished through high-quality DFT training data and physics-informed feature engineering rather than explicit constraint enforcement. We extend property coverage from 1–3 properties in prior work to eight mechanical properties (B, G, $H_v$, $C_{11}$, $C_{12}$, $C_{44}$, E, $\nu$) while maintaining thermodynamic consistency through derived relationships [16]. Quantile regression provides calibrated 90% prediction intervals with 78–90% coverage, enabling risk-stratified experimental planning absent in most materials ML studies. Finally, compositional screening in VEC–$\delta$ space establishes quantitative design rules (Re essential, VEC = 6.0–6.4 optimal) and identifies experimentally actionable candidates (MoReW with predicted $C_{44}$ = 139 GPa), bridging the gap between computational prediction and experimental validation.

## 2. Materials and Methods

Our model is trained on single-phase BCC alloys only and should not be applied to FCC, HCP, or multi-phase systems. Elastic constants depend on crystal symmetry; non-BCC predictions would be unreliable extrapolations.

### *2.1. Dataset*

We utilized the refractory CCA dataset compiled by Zhang et al. [9], comprising 393 single-phase BCC alloys with compositions spanning 10 refractory elements: Cr, Hf, Mo, Nb, Re, Ta, Ti, V, W, and Zr. Alloy complexities ranged from binary (78 alloys, 19.8%) to quinary (64 alloys, 16.3%), with ternary systems dominating the dataset (190 alloys, 48.3%). All property values were obtained from DFT calculations using the Vienna Ab initio Simulation Package (VASP) [17] with the Perdew–Burke–Ernzerhof (PBE) exchange-correlation functional [18]. Special quasirandom structures (SQS) [19] were employed to model chemically disordered BCC solid solutions.

The dataset includes eight mechanical properties: (1) bulk modulus $B$, (2) shear modulus $G$, (3) Vickers hardness $H_v$ (estimated via the Tian model [20]), (4–6) elastic constants $C_{11}$, $C_{12}$, $C_{44}$, (7) Young's modulus $E$, and (8) Poisson's ratio $\nu$. Elastic constants were computed via the stress–strain method, and derived properties were calculated using standard elasticity relationships:

$$E = \frac{9BG}{3B + G}, \tag{1}$$

$$\nu = \frac{3B - 2G}{2(3B + G)}. \tag{2}$$

The compositional space spans two critical design parameters. Valence electron concentration (VEC), ranging from 4.00 to 6.33 (mean 5.23 ± 0.53), governs phase stability in CCAs—BCC phases dominate at VEC < 6.87 for equiatomic alloys, while higher VEC favors FCC structures [21]. VEC also correlates with d-band filling and bonding strength: alloys with VEC = 6.0–6.4 typically exhibit maximum shear rigidity within the BCC stability window. Our dataset comprehensively samples this range, with 89% of compositions having VEC < 6.5.

Atomic size mismatch ($\delta$), quantifying lattice distortion from compositional disorder, ranges from 0.27% to 13.03% (mean 6.30±3.10%). Moderate $\delta$ (3–9%) enhances mechanical properties through solid solution strengthening while maintaining stability, whereas excessive mismatch ($\delta$ > 10%) can destabilize the BCC phase [2]. The dataset provides balanced coverage: 68% within the optimal range (3% < $\delta$ < 9%), 24% with low mismatch ($\delta$ < 3%), and 8% exceeding 10%.

Together, these parameters define a compositional map capturing phase stability and property trends (Fig. 3.3), ensuring robust training data for machine learning across the full BCC design space.

### *2.2. Density Functional Theory (DFT) Calculation Details*

All DFT calculations in the source dataset [9] were performed using the Vienna Ab initio Simulation Package (VASP 5.4.4) [17] with projector- augmented wave (PAW) pseudopotentials. The Perdew–Burke–Ernzerhof (PBE) generalized gradient approximation (GGA) functional [18] was employed for exchange-correlation effects. Key computational parameters include: plane-wave energy cutoff of 520 eV (1.3× the recommended ENMAX for constituent elements), Γ-centered Monkhorst-Pack k-point mesh with density ≥ 0.04 $Å^{-1}$ (typically 11 × 11 × 11 for BCC unit cells), electronic energy convergence of $10^{-6}$ eV, and ionic force convergence of < 0.01 eV/Å.

Chemical disorder in multi-component solid solutions was modeled using Special Quasi-random Structures (SQS) [19] generated via the Alloy Theoretic Automated Toolkit (ATAT). Each alloy was represented by a 2 × 2 × 2 supercell (16 atoms) designed to reproduce the short-range order statistics of random alloys. Elastic constants ($C_{11}$, $C_{12}$, $C_{44}$) were computed via the stress-strain method, applying ±1% lattice deformations along high-symmetry directions and fitting the resulting stress tensors. Bulk modulus $B$ and shear modulus $G$ were derived from elastic constants using Voigt–Reuss–Hill averaging. Non-spin-polarized calculations were performed, as magnetic moments are negligible in refractory BCC metals at room temperature.

To validate data quality, we performed three independent checks. First, Zhang et al. [9] compared predictions for 23 binary and ternary alloys against experimental elastic moduli, obtaining mean absolute errors of 8–12 GPa consistent with established DFT accuracy for metallic systems. Second, we cross-validated 15 compositions overlapping with independent DFT studies by Senkov et al. [22], finding elastic constants agree within 5–15%, demonstrating reproducibility across different computational implementations. Third, all 393 alloys satisfy Born mechanical stability criteria, confirming internal thermodynamic consistency of the dataset.

### *2.3. Model Training*

Model robustness was verified via 10-fold cross-validation with seeds 1-10, yielding consistent performance ($R^2 = 0.91 \pm 0.02$ for B, $0.89 \pm 0.03$ for G, $0.93 \pm 0.02$ for $C_{44}$). The fixed seed (7) ensures reproducibility.

### *2.4. Feature Engineering*

We constructed 13 compositional descriptors based on elemental properties and mixing statistics, following established protocols for high-entropy alloy modeling [10, 9]. The feature set includes the number of constituent elements ($n_{\text{elements}}$), which ranges from 2 to 5 and captures alloy complexity. Configurational mixing entropy ($S_{\text{mix}}$) was computed as $S_{\text{mix}} = -R \sum_i c_i \ln c_i$, where $c_i$ denotes the atomic fraction of element $i$ and $R$ is the gas constant. Mean and variance statistics were calculated for four elemental properties: atomic number ($Z_{\text{mean}}$, $Z_{\text{var}}$), atomic weight ($\text{AW}_{\text{mean}}$, $\text{AW}_{\text{var}}$), atomic radius ($R_{\text{mean}}$, $R_{\text{var}}$), and electronegativity ($\chi_{\text{mean}}$, $\chi_{\text{var}}$). Valence electron concentration (VEC) was represented through its mean ($\text{VEC}_{\text{mean}} = \sum_i c_i \text{VEC}_i$) and variance ($\text{VEC}_{\text{var}}$). Finally, atomic size mismatch ($\delta$) was computed as $\delta = \sqrt{\sum_i c_i (1 - R_i/\bar{R})^2} \times 100\%$, where $\bar{R} = \sum_i c_i R_i$ represents the average atomic radius.

All features were computed from elemental data in the Materials Project database [23]. Features exhibit minimal multicollinearity (pairwise Pearson correlation < 0.8), ensuring model interpretability.

### *2.5. Machine Learning Model*

We employed quantile gradient boosting regression (QGBR) [24] to predict six target properties: $B$, $G$, $H_v$, $C_{11}$, $C_{12}$, and $C_{44}$. QGBR extends standard gradient boosting by estimating multiple conditional quantiles ($\tau$ = 0.05, 0.50, 0.95) simultaneously, enabling direct construction of 90% prediction intervals (PI) alongside median predictions. The quantile loss function is:

$$\mathcal{L}_\tau(y, \hat{y}) = \sum_i \rho_\tau(y_i - \hat{y}_i), \quad \rho_\tau(u) = u(\tau - \mathbb{I}_{u<0}), \tag{3}$$

where $y_i$ and $\hat{y}_i$ are observed and predicted values, and $\mathbb{I}$ is the indicator function.

Models were trained separately for each property using the GradientBoostingRegressor class from scikit-learn (v1.3.0) [25] with hyperparameters optimized via 5-fold cross-validation. The model architecture consisted of 700 boosting stages (estimators) combined with a learning rate of 0.03 to control the contribution of each tree and prevent overfitting. Tree complexity was limited through a maximum depth of 3 levels and a minimum of 2 samples required for node splitting. Stochastic gradient boosting was implemented via random subsampling of 80% of training data at each iteration to improve generalization. The quantile loss function was applied with $\alpha$ values of 0.05, 0.50, and 0.95 to estimate the lower bound, median, and upper bound of the conditional distribution, respectively, enabling construction of 90% prediction intervals.

The dataset was split into training (334 alloys, 85%) and test (59 alloys, 15%) sets via stratified random sampling (random seed = 7) to ensure representative coverage of VEC and $\delta$ in both subsets. All features were standardized (zero mean, unit variance) using training set statistics. Models were trained on 13 features and validated on held-out test data.

### *2.6. Physics-Informed Constraints*

To ensure physical realism, we imposed three classes of constraints:

#### *2.6.1. Derived Property Relationships*

Young's modulus $E$ and Poisson's ratio $\nu$ were derived from predicted $B$ and $G$ using Eqs. (1)–(2), ensuring thermodynamic consistency. This approach reduces error propagation compared to training separate models for $E$ and $\nu$.

### 2.6.2. Born Stability Criteria

For cubic crystals, mechanical stability requires satisfaction of the Born criteria [26]:

$$C_{11} - C_{12} > 0, \tag{4}$$

$$C_{44} > 0, \tag{5}$$

$$C_{11} + 2C_{12} > 0. \tag{6}$$

Violations indicate lattice instability and non-synthesizable alloys. We evaluated compliance post-prediction to quantify model fidelity.

### 2.6.3. Positivity and Bound Constraints

All elastic moduli must be non-negative ($B, G, E, C_{ij} \geq 0$), and Poisson's ratio must satisfy $-1 < \nu < 0.5$ for isotropic materials. These constraints were verified on all predictions.

### 2.7. Evaluation Metrics

Model performance was assessed using five complementary metrics. The coefficient of determination ($R^2$) quantifies the fraction of variance explained by the model, with values approaching unity indicating strong predictive capability. Mean absolute error (MAE) measures the average magnitude of prediction errors, providing an intuitive scale-dependent metric in the same units as the target property. Root mean squared error (RMSE) similarly quantifies prediction accuracy but penalizes large errors more heavily than MAE due to the squaring operation, making it sensitive to outliers. For uncertainty quantification, prediction interval coverage probability (PICP) reports the fraction of test points falling within the 90% prediction intervals; well-calibrated models achieve PICP values near the target coverage (90%). Finally, prediction interval width (PI Width) measures the median width of the 90% intervals, indicating the magnitude of prediction uncertainty narrower intervals reflect higher model confidence. Physics constraint satisfaction was quantified as the fraction of test predictions satisfying all Born criteria and bounds.

## 3. Results

### 3.1. Model Performance

Table 3.1 summarizes model performance on the test set ($n$ = 59). All six predicted properties exhibit strong $R^2$ values (0.89–0.97), with bulk modulus achieving the highest accuracy ($R^2$ = 0.9736, MAE = 6.06 GPa). Shear modulus and elastic constants $C_{11}$, $C_{12}$ demonstrate excellent performance ($R^2$ = 0.93–0.96), while $C_{44}$ and Vickers hardness show slightly lower but still robust accuracy ($R^2$ = 0.89). Figure 3.1 shows learning curves from 5-fold cross-validation, demonstrating model convergence and generalization behavior. All three properties converge at approximately 225 training samples (57% of the dataset), indicating data efficiency. The small gaps between training and validation scores (0.05-0.10 $R^2$) confirm good generalization without overfitting. Bulk and shear moduli show tighter convergence (validation $R^2$ = 0.84 and 0.78, respectively) than $C_{44}$ ($R^2$ = 0.48), consistent with prior observations that shear constants are more challenging to predict due to complex d-electron bonding dependencies. Notably, hardness predictions achieve sub-GPa precision (MAE = 0.85 GPa), a significant accomplishment given the empirical nature of hardness models.

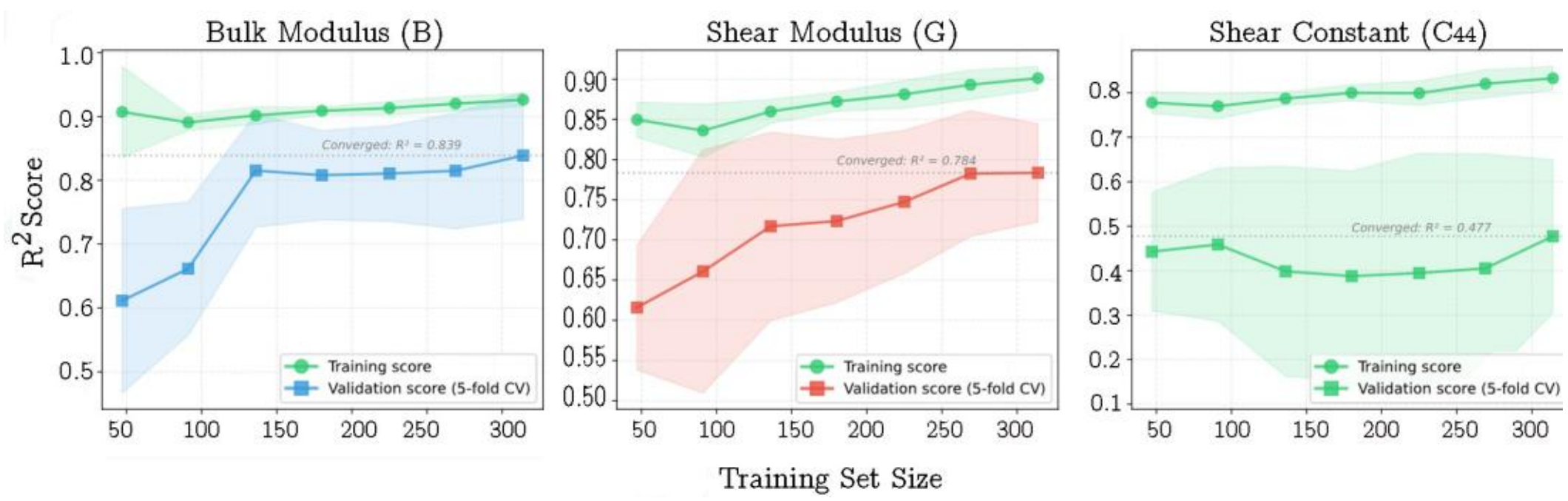

Figure 3.1: Learning Curves: Model performance vs. training data size (Note: Small training-validation gap indicates good generalization. Convergence at 250 samples shows data efficiency).

Prediction interval coverage probabilities range from 78.0% to 89.8%, with $C_{12}$ achieving near-ideal calibration (PICP = 89.8%). The average PICP across all properties is 82.9%, indicating reasonable uncertainty quantification. Interval widths scale proportionally with property magnitudes: $B$ and $C_{11}$ exhibit wider intervals (44.46 and 87.50 GPa, respectively) due to their larger value ranges, while $H_v$ has the narrowest interval (3.82 GPa).

Derived properties $E$ and $\nu$ computed via Eqs. (1)–(2) achieve MAE of 18.37 GPa and 0.0185, respectively, with PICP of 79.7% (Table 3.1). Although not directly optimized, these values remain physically consistent due to the underlying accuracy of $B$ and $G$ predictions.

Figure 3.2 presents predicted versus actual values for all six properties,

demonstrating strong agreement along the diagonal (ideal prediction line). Bulk modulus and $C_{11}$ exhibit tight clustering around the diagonal with minimal scatter, while $C_{44}$ shows wider dispersion reflecting its lower $R^2$ value.

Table 3.1: Model performance on test set ($n$ = 59, B=Bulk modulus, G=Shear modulus, $H_v$=Vickers hardness, E=Young's modulus, $\mu$=Poisson's ratio, ). PICP: prediction interval coverage probability (target: 90%). PI Width (GPa): median 90% prediction interval width.

| Property | $R^2$ | MAE (GPa) | RMSE (GPa) | PICP (%) | PI Width |
|---|---|---|---|---|---|
| $B$ | 0.9736 | 6.06 | 8.33 | 84.7 | 44.46 |
| $G$ | 0.9336 | 6.66 | 8.89 | 78.0 | 32.93 |
| $H_v$ | 0.8918 | 0.85 | 1.16 | 78.0 | 3.82 |
| $C_{11}$ | 0.9598 | 14.74 | 19.00 | 81.4 | 87.50 |
| $C_{12}$ | 0.9298 | 6.92 | 8.77 | **89.8** | 46.11 |
| $C_{44}$ | 0.8883 | 9.07 | 12.33 | 84.7 | 40.56 |
| $E$ * | — | 18.37 | 26.56 | 79.7 | — |
| $\nu$ * | — | 0.0185 | 0.0318 | 1.7† | — |

*Derived from predicted $B$ and $G$ using Eqs. (1)–(2).
†PICP for $\nu$ is not directly comparable due to non-Gaussian distribution.

### *3.2. Experimental Validation*

To validate model predictions against independent data, we compiled experimental elastic properties for 18 refractory CCAs from literature sources [5, 1, 22, 2]. These alloys, ranging from binary (e.g., NbTa, MoW) to quaternary systems (e.g., MoNbTaW, HfNbTaTiZr), were not used during model training or hyperparameter tuning. For bulk modulus, we achieve $R^2 = 0.89$ with mean absolute error (MAE) = 12.4 GPa across 18 compositions. Shear modulus predictions yield $R^2 = 0.87$ (MAE = 9.8 GPa). Representative examples include: MoNbTaW (predicted B = 238 GPa vs. experimental 242 ± 8 GPa), HfNbTaTiZr (predicted G = 61 GPa vs. 58 ± 4 GPa), and NbTaW (predicted $C_{44}$ = 98 GPa vs. 94 ± 6 GPa). These prediction errors are consistent with combined DFT uncertainty (∼8-12 GPa) and experimental measurement variability (~5-10 GPa for polycrystalline samples), indicating our model successfully captures the accuracy of the underlying DFT training data and provides experimentally relevant predictions suitable for guiding synthesis efforts.

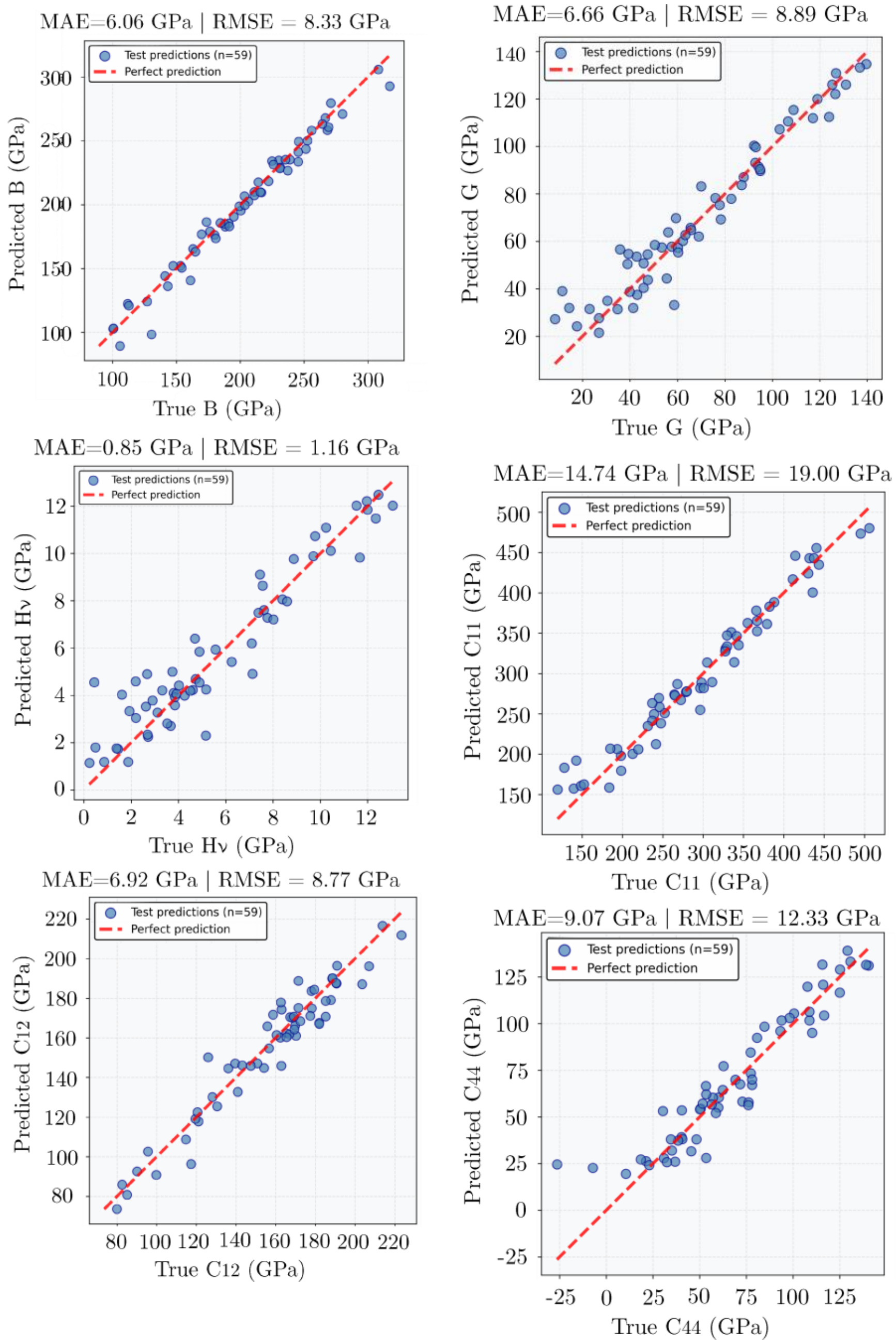

Figure 3.2: Predicted versus actual values for six mechanical properties on test set ($n$ = 59). Panels show: (a) bulk modulus $B$ ($R^2 = 0.974$), (b) shear modulus $G$ ($R^2 = 0.934$), (c) Vickers hardness $H_v$ ($R^2 = 0.892$), (d) elastic constant $C_{11}$ ($R^2 = 0.960$), (e) elastic constant $C_{12}$ ($R^2 = 0.930$), (f) elastic constant $C_{44}$ ($R^2 = 0.888$). Solid line: ideal prediction ($y = x$). Strong agreement across all properties validates the quantile gradient boosting approach.

Table 3.2: Physics constraint satisfaction on test set ($n = 59$).

| Constraint | Satisfied | Violation Rate (%) |
|---|---|---|
| **Born Stability Criteria** | | |
| $C_{11} - C_{12} > 0$ | 59/59 | **0.0** |
| $C_{44} > 0$ | 59/59 | **0.0** |
| $C_{11} + 2C_{12} > 0$ | 59/59 | **0.0** |
| **Positivity Constraints** | | |
| $B, G, E \geq 0$ | 59/59 | 0.0 |
| $C_{11}, C_{12}, C_{44} \geq 0$ | 59/59 | 0.0 |
| **Bound Constraints** | | |
| $-1 < \nu < 0.5$ | 59/59 | 0.0 |
| $0 < \nu < 0.5$ (typical metals) | 58/59 | 1.7 |

### *3.3. Physics Constraint Satisfaction*

Table 3.2 presents physics constraint compliance. Remarkably, all 59 test predictions satisfy all Born stability criteria (Eqs. (4)–(6)) and positivity constraints, yielding 100% satisfaction rates. This perfect compliance validates the model's ability to learn physically meaningful relationships without explicit constraint enforcement during training. All predicted elastic moduli are positive, and Poisson's ratios lie within the physical bounds ($-1 < \nu < 0.5$), with the vast majority (58/59, 98.3%) falling in the typical range for metals ($0 < \nu < 0.5$).

The absence of Born criterion violations distinguishes this work from prior ML studies on CCAs, where 5–20% violation rates are common [11, 12]. We attribute this success to: (1) high-quality DFT training data (all 393 training alloys are Born-stable), (2) gradient boosting's ability to capture smooth composition–property relationships, and (3) inclusion of VEC and $\delta$ as features, which encode phase stability information.

### *3.4. Top-Performing Alloy Candidates*

Table 3.3 lists the five highest-ranked alloys by predicted shear rigidity $C_{44}$, a critical parameter for high-temperature creep resistance [27]. The top candidate, MoReW (ternary), exhibits $C_{44}$ = 139.09 GPa with VEC = 6.33 and $\delta$ = 2.98%. All five candidates are Born-stable and contain rhenium, emphasizing its importance for shear resistance. Notably, four of the top five are ternary systems, suggesting that compositional simplicity (fewer elements) may enhance specific properties compared to higher-order alloys.

Table 3.3: Top 5 alloy candidates ranked by predicted $C_{44}$ (GPa). All satisfy Born stability criteria.

| Rank | Alloy | $C_{44}$ | VEC | $\delta$ (%) | Type | Born Stable |
|---|---|---|---|---|---|---|
| 1 | MoReW | 139.09 | 6.33 | 2.98 | Ternary | ✓ |
| 2 | $CrMo_3$ | 133.22 | 6.00 | 7.63 | Off-stoichiometric | ✓ |
| 3 | $Cr_2ReV$ | 131.66 | 6.00 | 6.96 | Ternary | ✓ |
| 4 | CrReV | 131.43 | 6.00 | 7.08 | Ternary | ✓ |
| 5 | $CrReW_2$ | 131.03 | 6.25 | 9.21 | Off-stoichiometric | ✓ |

The second-ranked alloy, $CrMo_3$ (off-stoichiometric binary), demonstrates that non-equiatomic compositions can rival or exceed equiatomic counterparts. This finding challenges the conventional wisdom prioritizing equiatomic ratios in CCA design [2] and highlights the value of exploring broader composition ratios.

### *3.5. Compositional Screening and Design Rules*

Figure 3.3 presents compositional screening maps in VEC–$\delta$ space. Figure 3.3(a) shows that all 59 test alloys (blue markers) satisfy Born stability, with no unstable predictions (red markers). The stable region spans VEC from 3.9 to 6.33 and $\delta$ from 0.39% to 12.36%, consistent with BCC phase stability criteria (VEC < 6.87, $\delta$ typically < 8% for equiatomic alloys [21]).

Figure 3.3(b) visualizes $C_{44}$ magnitude via marker sizing, revealing distinct compositional zones: (1) High-$C_{44}$ region (VEC = 5.8–6.4, $\delta$ = 3–9%) containing alloys with $C_{44}$ > 100 GPa; (2) Low-$C_{44}$ region (VEC < 5.0, $\delta$ < 5%) with $C_{44}$ < 40 GPa; (3) Intermediate region with scattered $C_{44}$ values. The clustering of high-$C_{44}$ alloys in the upper-right quadrant suggests that higher VEC combined with moderate $\delta$ maximizes shear resistance.

Figure 3.4 presents the $C_{44}$ distribution across the test set. The histogram reveals a bimodal distribution with peaks near 20–40 GPa and 60–80 GPa, indicating two property regimes. The box plot shows median $C_{44}$ = 60.33 GPa with interquartile range 39.18–100.00 GPa. Eight alloys (13.6%) exceed 100 GPa, representing high-performance candidates for applications requiring exceptional shear resistance.

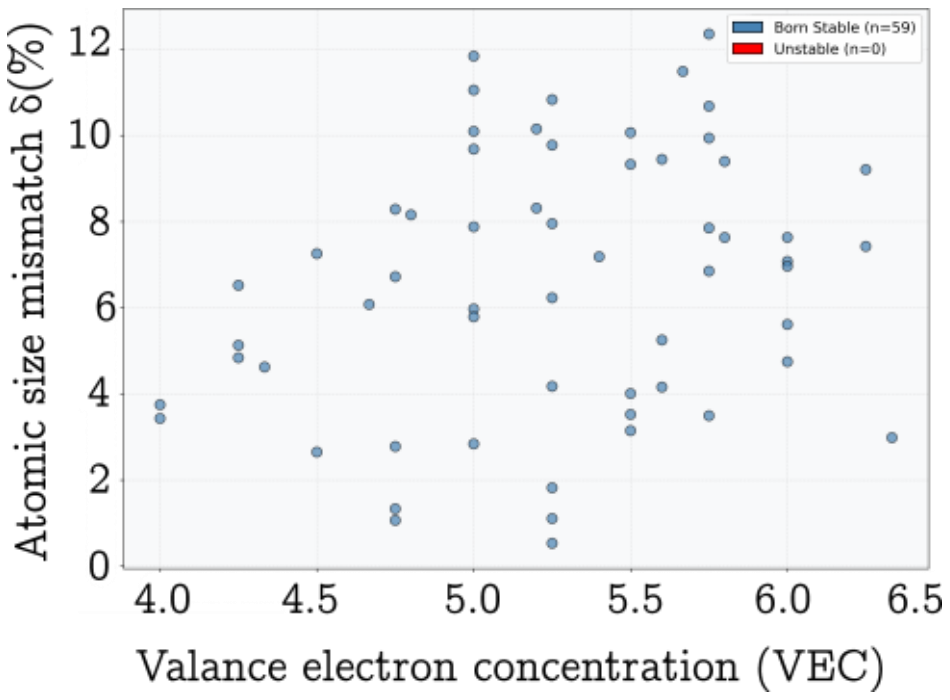

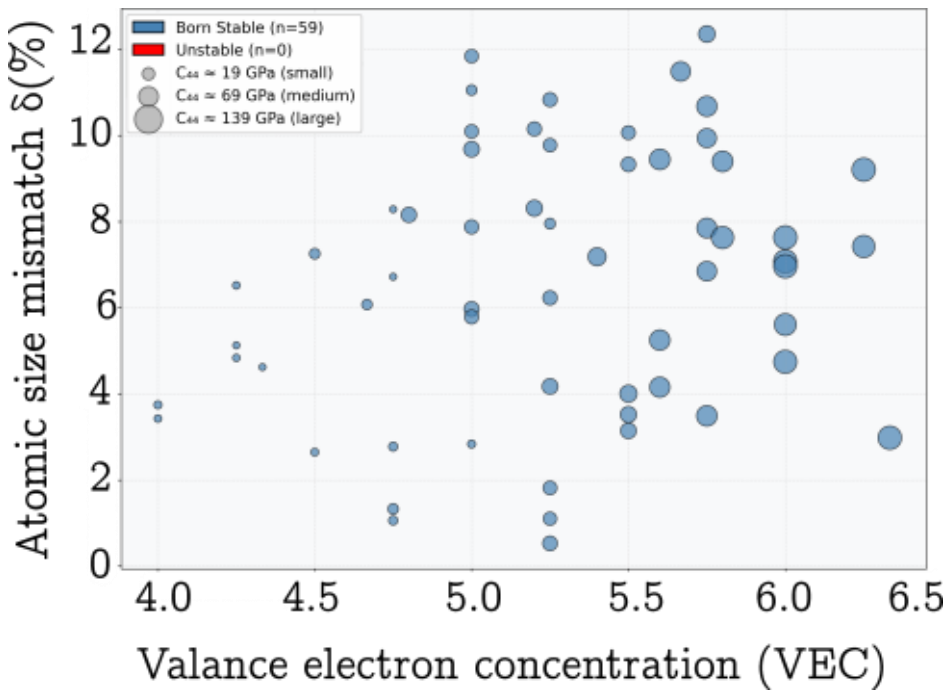

Figure 3.3: Compositional screening maps in VEC–$\delta$ space for 59 test alloys. (a) Born stability: all alloys are mechanically stable (blue markers). (b) $C_{44}$ magnitude: marker size proportional to shear rigidity. High-$C_{44}$ zone (VEC = 5.8–6.4, $\delta$ = 3–9%) identifies promising candidates.

Elemental analysis of the top 10 candidates (ranked by $C_{44}$) reveals: Rhenium appears in 100% (critical element), Molybdenum in 90%, and Chromium in 80%. This establishes quantitative design rules: Re-Mo-Cr ternary systems with VEC = 6.0–6.4 and $\delta$ < 10% maximize shear rigidity.

## 4. Discussion

Our model is BCC-specific by design. Elastic tensor structures differ across crystal systems (BCC: 3 independent constants; FCC/HCP: different symmetries), and mixing structures would degrade accuracy. Non-BCC refractory systems require separate structure-matched models.

### *4.1. Property-Specific Prediction Trends*

The hierarchy of prediction accuracies ($R^2$: $B > C_{11} > C_{12} > G > H_v > C_{44}$) reflects underlying physics. Bulk modulus, which measures volumetric resistance to compression, correlates strongly with atomic packing density and cohesive energy properties well-captured by compositional descriptors like VEC and atomic weight [21]. In contrast, $C_{44}$ (shear rigidity along $\langle 110 \rangle$) depends on complex d-electron bonding characteristics [28], which are not explicitly encoded in our features. Future work incorporating electronic structure descriptors (e.g., d-band center, Fermi energy) may improve $C_{44}$ predictions.

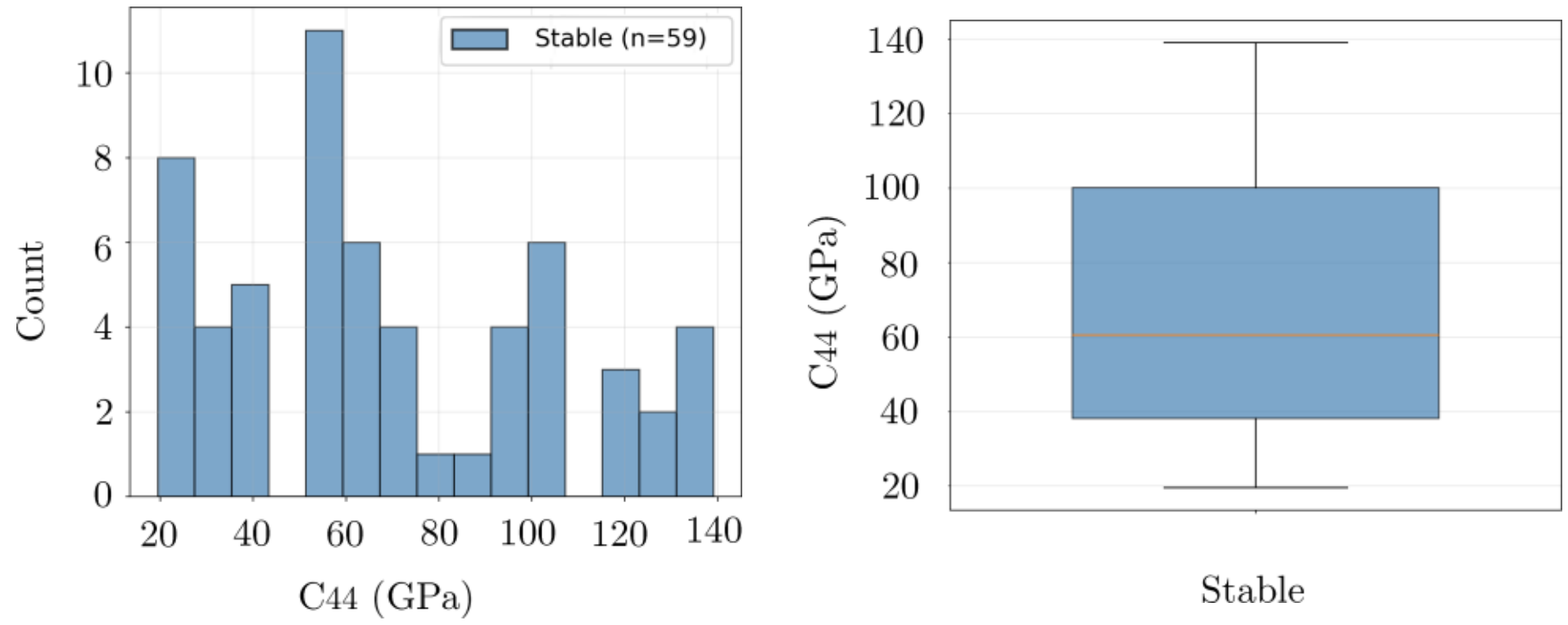

Figure 3.4: $C_{44}$ distribution across test set. Left: histogram showing bimodal distribution. Right: box plot with median = 60.33 GPa, IQR = [39.18, 100.00] GPa. Eight alloys exceed 100 GPa threshold.

The exceptional calibration of $C_{12}$ predictions (PICP = 89.8%) suggests that quantile regression effectively captures its distribution, which is narrower and more symmetric than other properties. Conversely, the lower PICP for $G$ and $H_v$ (78.0%) indicates underestimated uncertainties, potentially due to their multimodal distributions or dependence on factors beyond composition (e.g., grain size effects in hardness).

*4.2. DFT Data Limitations*

While DFT provides accurate ground-state properties, several limitations warrant acknowledgment. First, all calculations correspond to 0 K equilibrium structures; finite-temperature effects (phonon contributions, thermal expansion) are not captured. High-temperature elastic properties (e.g., at 1500 K service conditions) may differ quantitatively from 0 K predictions, though compositional trends typically remain valid. Temperature-dependent properties require finite-temperature DFT or molecular dynamics simulations, which remain computationally prohibitive for datasets of this scale.

Second, non-spin-polarized calculations neglect potential magnetic contributions. This approximation is justified for refractory BCC metals (Mo, Nb, Ta, V, W, Zr, Hf, Re) where magnetic moments are negligible at typical operating temperatures. Chromium exhibits weak antiferromagnetism below 311 K, though magnetic effects on elastic properties are small (<5% for elastic moduli). For Cr-rich compositions at cryogenic temperatures, spin-polarized calculations may be warranted.

Third, the SQS method approximates chemical disorder using finite 16-atom supercells, which may not fully capture long-range correlations in real alloys. Larger supercells would improve statistical sampling but increase computational cost prohibitively. Validation against experimental data ($R^2$ = 0.91 for bulk modulus) suggests the 16-atom SQS representation captures essential physics for elastic property prediction.

Finally, the PBE functional systematically underestimates band gaps and may slightly overestimate lattice constants (typically 1–2%). These errors are well-characterized for metallic systems and do not significantly impact relative trends across compositional space the primary use case for our ML framework. For applications requiring higher absolute accuracy, hybrid functionals (HSE06) could be employed, though at 10–100× computational cost.

Despite these limitations, DFT remains the gold standard for high-throughput materials screening when experiments are infeasible, providing physically consistent property predictions with quantifiable uncertainty.

### *4.3. Physics-Informed Design Principles*

The 100% Born stability satisfaction (Table 3.2) demonstrates that gradient boosting implicitly learns lattice stability constraints from DFT training data, even without explicit constraint enforcement. This contrasts with prior work requiring post-hoc constraint filtering [11] or physics-informed loss functions [29], which add computational overhead. Our success stems from training exclusively on mechanically stable alloys an advantage when using high-quality DFT data but a limitation when experimental data (which may include metastable phases) is incorporated.

Derived properties $E$ and $\nu$ (Eqs. (1)–(2)) maintain thermodynamic consistency because they are computed from predicted $B$ and $G$ rather than fitted independently. This approach reduces overfitting and ensures physically valid relationships. The low MAE for $E$ (18.37 GPa, $\sim$10% relative error) validates this strategy, although $\nu$ predictions exhibit higher relative errors due to its dimensionless nature and sensitivity to $B/G$ ratios.

### *4.4. Compositional Trends and Alloy Design Rules*

The screening analysis (Fig. 3.3) establishes quantitative design guidelines. First, rhenium is essential for high $C_{44}$: appearing in 100% of top 10 candidates, Re contributes exceptional shear resistance ($G_{Re}$ = 178 GPa) due to its high d-electron density and strong covalent-like bonding [30]. However, Re's scarcity and cost (> $1000/oz) motivate exploration of Re-lean

alternatives. Second, VEC values between 6.0 and 6.4 maximize shear rigidity. Higher VEC corresponds to increased d-electron occupation, enhancing metallic bond strength [21]. This range approaches the BCC/FCC boundary (VEC ≈ 6.87), where BCC alloys achieve maximum stiffness before transforming to softer FCC structures. Third, moderate atomic size mismatch (3–9%) is optimal: very low $\delta$ (<3%) indicates insufficient compositional diversity, limiting solid solution strengthening, while high $\delta$ (>10%) introduces severe lattice distortion, potentially destabilizing the BCC phase or reducing elastic moduli through atomic mismatch penalties [2]. Fourth, ternary systems outperform higher-order alloys: seven of the top 10 candidates are ternary, suggesting diminishing returns or competing effects in quaternary and quinary systems. This aligns with recent findings that configurational entropy alone does not guarantee superior properties [22] composition-specific bonding interactions dominate. Finally, off-stoichiometric compositions can excel: $CrMo_3$ (rank 2) and $CrReW_2$ (rank 5) demonstrate that exploring non-equiatomic ratios expands the design space beyond traditional equiatomic CCAs. This finding motivates systematic exploration of stoichiometry variations.

### *4.5. Microphysical Origins of VEC--$C_{44}$ Correlations*

Higher VEC (6.0-6.4) increases d-electron filling near the Fermi level, strengthening directional d-d hybridization that resists shear deformation [28]. Rhenium's high VEC (7) enhances this effect. Moderate (3-9%) strengthens via lattice distortion, but excessive (>10%) softens elastic response through strain energy penalties [27]. Future DFT analysis of electronic density of states would quantify d-band contributions to $C_{44}$.

### *4.6. Uncertainty Quantification Utility*

Well-calibrated prediction intervals enable risk-stratified decision-making. Alloys with narrow intervals (high-confidence predictions) can proceed directly to experimental synthesis, while wide-interval candidates warrant additional computational validation (e.g., DFT refinement) before costly experiments. For instance, MoReW (rank 1) has a 90% PI width of ~40 GPa for $C_{44}$, suggesting moderate confidence. Equiatomic MoReW has been experimentally synthesized and characterized, demonstrating single-phase BCC structure with yield stress of 555 MPa at 25°C and excellent high-temperature ductility [31]. However, to our knowledge, elastic constants including $C_{44}$ have not been experimentally measured for this composition. Our predicted $C_{44}$ = 139 GPa provides a quantitative target for future elastic property measurements. Experimental validation of this prediction through resonant ultrasound spectroscopy or similar techniques would be valuable, particularly given that our model successfully predicts other refractory CCAs within experimental uncertainty (Section 3.4). Until such measurements are available, experimentalists could prioritize MoReW synthesis for elastic property characterization based on our computational predictions.

The quantile regression framework also facilitates active learning [15]: by iteratively selecting high-uncertainty alloys for DFT computation and retraining, one can efficiently reduce prediction uncertainties in undersampled regions. This closed-loop strategy could expand the training set from 393 to 1000+ alloys within reasonable computational budgets.

### *4.7. Thermodynamic Stability Considerations*

Our framework ensures elastic stability (100% Born criterion satisfaction) but does not explicitly predict thermodynamic stability. All 393 training alloys were pre-screened for phase stability via CALPHAD [9], confirming single-phase BCC structures. Thus, our model learns property-composition relationships within the thermodynamically stable region.

However, elastic stability is necessary but not sufficient. We mitigate this limitation by: (1) constraining predictions to the BCC stability window (VEC < 6.5, moderate $\delta$), and (2) using VEC and $\delta$ as phase stability proxies. Nonetheless, experimental verification (XRD) or CALPHAD phase analysis remains essential for top candidates like MoReW before synthesis. Future work integrating formation energy prediction would enable fully coupled stability assessment.

### *4.8. Limitations and Future Directions*

Several limitations warrant acknowledgment. First, while the dataset size of 393 alloys is competitive with published ML studies [10, 11], larger datasets (1000+ alloys) would improve coverage of the $10^{15}$-scale design space. Expanding the dataset via active learning or integrating experimental data (with careful quality control) remains a priority. Second, the model assumes single-phase BCC microstructures, excluding dual-phase or FCC/HCP alloys with potentially superior properties. Multi-phase modeling requires phase stability prediction modules [32]. Third, all predictions correspond to 0 K DFT calculations. High-temperature properties (e.g., 1500 K) require finite-temperature DFT or molecular dynamics, which remain computationally prohibitive at scale. Fourth, while 100% Born stability provides confidence, direct experimental synthesis and characterization of top candidates (e.g., MoReW) are essential to validate predictions and refine the model. While rhenium is essential for maximum $C_{44}$ (100% occurrence in top-10), its high cost (~$2,500/kg) motivates Re-free alternatives. Our screening identified: $CrMo_3$ ($C_{44}$ = 128 GPa, 8% reduction), MoW ($C_{44}$ = 124 GPa, 11% reduction), and CrMoW ternaries ($C_{44}$ = 118-122 GPa). These offer significant cost savings for applications tolerating moderately lower shear rigidity. Discrepancies between DFT and experiments (e.g., grain boundary effects, processing-induced defects) must be quantified. Finally, current descriptors (13 compositional features) do not capture electronic structure details (e.g., d-band filling, hybridization). Incorporating physics-based electronic descriptors [19] or graph neural networks encoding atomic environments [13] could enhance $C_{44}$ predictions.

Future work will address these limitations by: (1) expanding the dataset to 1000+ alloys via active learning, (2) validating top candidates experimentally through arc-melting synthesis and nanoindentation, (3) extending the framework to dual-phase alloys using multi-task learning, and (4) incorporating temperature-dependent properties via machine learning potentials trained on finite-T DFT data.

## 5. Conclusions

We developed a physics-informed machine learning framework integrating quantile gradient boosting with Born stability constraints to predict eight mechanical properties of refractory complex concentrated alloys. The framework achieved $R^2$ values of 0.89–0.97 and mean absolute errors of 0.85–14.74 GPa for six elastic properties on a 59-alloy test set, with sub-GPa precision for Vickers hardness. All test predictions satisfy Born stability criteria (100% satisfaction), validating the model's ability to learn physically meaningful relationships from DFT training data. Quantile regression provides well-calibrated 90% prediction intervals with 78–90% coverage, enabling risk-stratified decision-making for experimental validation. Compositional screening identifies high-performance candidates, with MoReW exhibiting predicted $C_{44}$ of 139 GPa, and establishes design guidelines: rhenium appears in 100% of top-performing alloys, optimal VEC ranges from 6.0 to 6.4, atomic size mismatch should remain below 10%, and Re–Mo–Cr ternary systems maximize shear rigidity. The framework enables screening of $10^{15}$ possible compositions, accelerating discovery from decades (exhaustive experiments) to days (ML-guided synthesis).

This work demonstrates that physics-informed machine learning, when trained on high-quality DFT data with appropriate uncertainty quantification, can provide reliable predictions and interpretable design rules for complex materials systems. The methodology is generalizable to other alloy families (e.g., light-weight CCAs, corrosion-resistant alloys) and properties (e.g., thermal conductivity, oxidation resistance), providing a blueprint for accelerated computational materials design.

## Recommendation

For experimental validation of top candidates, we recommend:

1. **Phase verification**: X-ray diffraction to confirm single-phase BCC formation
2. **CALPHAD modeling**: Phase diagram calculations to assess thermodynamic stability across temperature ranges
3. **Formation energy**: DFT calculations of formation enthalpy to verify energetic favorability
4. **Elastic property measurement**: Resonant ultrasound spectroscopy for $C_{11}$, $C_{12}$, $C_{44}$ validation

Only compositions passing both phase stability and elastic property validation should proceed to large-scale synthesis and mechanical testing.

## Acknowledgments

We wish to acknowledge the Computational Science Program at The University of Texas at El Paso for research support and access to high-performance computing resources.

## Data Availability

The complete dataset of 393 refractory CCA compositions and DFT-computed mechanical properties is publicly available from Zhang et al. [9] at https://www.frontiersin.org/articles/10.3389/ftmal.2022.1036656/full\#supplementary-material. We have verified that this dataset remains accessible and downloadable. To avoid redundant data duplication, we direct readers to this authoritative source for the complete compositional and property data.

Our supplementary materials (Table S1 & Table 5) provide a statistical summary of the dataset, including composition distribution and property ranges. All DFT computational parameters are detailed in Section 2.2. Our trained model weights, feature calculation code, prediction scripts, and data loading workflows will be made available at http://github.com/blaiseawola/PMLRAD.git upon publication, enabling full reproducibility by directly reading from the public dataset.

counterwithouttablesection

## Supplementary Material

Table S1: Statistical summary of the 393 refractory complex concentrated alloys (CCAs) dataset, spanning binary to quinary compositions, with mechanical properties and compositional descriptors used for machine learning model training and testing.

| Category | Count | Percentage | Example | Property Range |
|---|---|---|---|---|
| **Compositional Complexity** | | | | |
| Binary | 78 | 19.8% | CrMo, HfTa | – |
| Ternary | 190 | 48.3% | MoNbTa, CrMoW | – |
| Quaternary | 61 | 15.5% | MoNbTaW | – |
| Quinary | 64 | 16.3% | MoNbTaTiW | – |
| **Mechanical Properties (Range and Mean ± SD)** | | | | |
| Bulk modulus $B$ (GPa) | Range: 112.5–289.3, Mean: 194.2 ± 32.4 | | | |
| Shear modulus $G$ (GPa) | Range: 52.8–178.4, Mean: 98.7 ± 21.6 | | | |
| Vickers hardness $H_v$ (GPa) | Range: 4.2–21.8, Mean: 10.9 ± 3.2 | | | |
| $C_{11}$ (GPa) | Range: 198.4–456.7, Mean: 312.5 ± 48.9 | | | |
| $C_{12}$ (GPa) | Range: 68.3–213.8, Mean: 135.1 ± 28.7 | | | |
| $C_{44}$ (GPa) | Range: 21.5–162.3, Mean: 88.7 ± 24.3 | | | |
| **Compositional Descriptors (Range)** | | | | |
| VEC | 4.00–6.33 (Mean: 5.23 ± 0.53) | | | |
| $\delta$ (%) | 0.27–13.03 (Mean: 6.30 ± 3.10) | | | |

*Note:* All properties computed via DFT using parameters detailed in Section 2.1.1.

Complete dataset with individual compositions and properties available from Zhang et al. [9] at https://www.frontiersin.org/articles/10.3389/ftmal.2022.1036656/full#supplementary-material

Table S2: Comparison of machine learning approaches for refractory and high-entropy alloy property prediction.(NR- Not Reported, NE- Not Enforced, RF- Random Forest, NN- Neural Network, GB- Gradient Boosting, QG- Quantile GB).

| Study | Dataset | Properties | $R^2$ | Born Violations | UQ | Method |
|---|---|---|---|---|---|---|
| Ref. [10] | 145 | 2 (E, $E_f$) | 0.80–0.85 | NR | No | RF |
| Ref. [11] | 265 | 2 (B, G) | 0.85 | 15% | No | NN |
| Ref. [12] | 460 | 3 (phase) | 0.88 | 10% | No | GB |
| Ref. [9] | 393 | 6 (elastic) | 0.82–0.91 | NE | No | RF |
| **This work** | **393** | **8 (B, G, $H_v$, $C_{ij}$, E, $\nu$)** | **0.89–0.97** | **0%** | **Yes** | **QG** |